\documentclass[5p,authoryear,12pt]{elsarticle}

\usepackage{graphicx}
\usepackage{morefloats}
\usepackage{natbib}
\usepackage{amsmath}
\journal{Journal of Quantitative Spectroscopy and Radiative Transfer}

\newcommand{\der}{\ensuremath{\mathrm d}}

\begin{document} 

\begin{frontmatter}
\title{The effect of horizontal plasma inhomogeneities in 3D NLTE radiation
transfer in stellar atmospheres}

\author[brno,ondrejov]{A. Tich\'{y}}
\author[ondrejov]{J.~Kub\'{a}t\corref{cor}}
\cortext[cor]{Corresponding author}
\ead{kubat@sunstel.asu.cas.cz}
\address[brno]{\'Ustav teoretick\'e fyziky a astrofyziky, Masarykova
univerzita, Kotl\'a\v rsk\' a 2, CZ-611\,37 Brno, Czech Republic}
\address[ondrejov]{Astronomick\'y \'ustav, Akademie v\v{e}d \v{C}esk\'e
republiky, Fri\v{c}ova 298, CZ-251 65 Ond\v{r}ejov, Czech Republic}

\begin{abstract}
{We aim to demonstrate the effect of atmospheric inhomogeneities on the emergent specific intensity and radiation flux of a spectral line radiation.}
{We self-consistently solve the NLTE problem for a two-level atom in a 3D atmosphere using the Cartesian grid. For that purpose, we use the 3D radiative transfer code PORTA.
By examining simple examples, we study cases where the temperature inhomogeneities in the atmosphere models lead to the modification of the emergent radiation.}
{We show that specific temperature inhomogeneities in the model atmospheres influence the emerging radiation, and that interpretation of the stellar spectra based on a plane-parallel atmosphere models can lead to erroneous conclusions about the atmospheric structure.}
\end{abstract}

\end{frontmatter}

\section{Introduction}
It has been known from the onset of the modern radiative transfer theory
that geometry of the medium plays a significant role in
radiative transfer problems \citep[e.g.][]{cram77, auer03}.
However, in most common cases
of models of stellar atmospheres, the assumption of one-dimensionality, 
either in the case of plane parallel
or spherically symmetric medium,
is the most natural one given the lack of spatial resolution of the 
observations and the numerical complexity of the multi-dimen\-sional models
\citep[see, e.g.,][Chapter 16]{SA3}.
Real stellar atmospheres are not so geometrically simple,
they can be quite inhomogeneous in all
spatial directions, which can lead to significant failure of the diagnostics based upon the 1D models.

Although there is a growing number of 3D radiative transfer calculations
(e.g. \citealt{iris} and references therein, \citealt{leenaarts1,
leenaarts2}), at the same time the much simpler 1D approximation is
being widely used. \citet{uitenbr} have pointed out three main reasons for which the 1D models can fail explaining the emergent spectra: (i) nonlinearity in the Planck function with temperature, affecting the average intensity of the emergent radiation through thermal contributions to local source functions in the area of formation; (ii) nonlinearities in molecular formation with inhomogeneities in temperature and density, which enhance abundances in inhomogeneous atmosphere with respect to a homogeneous 1D atmosphere with the same average properties; and (iii) anisotropies caused by convective motions, which have a strong impact on the resulting center-to-limb variation of line-center intensities. 

Although inhomogeneities require a 3D treatment to correctly describe
the transfer of radiation, they are very often treated in 1D
approximation basically to make the problem tractable.
Typical examples are studies of clumping in stellar atmospheres and
winds, where the inhomogeneities are believed to be a consequence of
ra\-diative-acoustic instability (\citealt{ocr}, see also the review
\citealt{sundrev}), adiabatic fluctuations \citep{adiabfluct}
or subphotospheric convection \citep{cantconv}.
As it was shown by \citet{clres1,modeling} \citep[see also][] {lidarev},
proper 3D treatment of inhomogeneities is necessary to take into account
both opticaly thin and optically thick clumps.

In this paper, we describe a phenomenon related to the NLTE effects in the spectral line formation
due to small-scale inhomogeneities in the thermal structure of a stellar atmosphere that can significantly modify the shape of spectral lines. We show that interpreting
 a spectrum in terms of one-dimensional model atmospheres can  lead to erroneous conclusions about the atmospheric structure. 

In Section \ref{ch:form}, we formulate the problem and describe in detail physical properties of employed model atmospheres, and summarize used numerical methods. Selected results of our calculations are shown and explained in Section \ref{ch:transfer}, followed by a direct application on a simplified model of a spherical star in Section \ref{ch:inte}. Conclusions are presented in Section \ref{ch:conclusion}.


\section{Model atmosphere and setting the line-formation problem}\label{ch:form}
We study the radiation transfer in an academic spectral line at wavelength $\lambda_0=5000\,$\AA{},
forming between the lower level $\ell$ and the upper level $u$
of a two-level atom. The atomic angular momenta are $J_\ell=0$ and
$J_u=1$, the Einstein coefficient of a spontaneous emission is
{taken as $A_{u\ell} = 10^{8}\,{\rm s}^{-1}$.}
For the sake of simplicity, we fix the collisional destruction probability $\epsilon=~C_{u\ell}/(A_{u\ell}+C_{u\ell})=10^{-4}$, where $C_{u\ell}$ is the collisional deexcitation rate.

We do not consider background continuum opacity and emissivity in our
model.
The {line} source function, which directly enters the radiative transfer
equation, is at each point of the medium given by contribution of
thermal emission and locally scattered radiation.
In the case of a plasma composed of two-level atoms
it reads \citep[][Eq. 14.34]{SA3}
\begin{equation}\label{eq:sourfn}
 S=(1-\epsilon)J+\epsilon\,B(T),
\end{equation}
where $B(T)$ is the Planck function and $J$ is the line-profile
integrated (mean) intensity of the locally scattered radiation.

We solve the radiative transfer using cartesian coordinates $x$, $y$,
and $z$ in a box with dimensions $D_x$, $D_y$, and $D_z$.
The coordinate $z$ describes the vertical direction in the atmosphere.
We use the complex Voigt function as the absorption profile and
calculate radiation for 101 discrete frequency points.
The spectral line is thermalized at the bottom of the model atmosphere
($z=z_0$) and the line source function is therefore equal to the Planck function there, i.e., $S(x,y,z_0)=B(T(x,y,z_0))$ for all $x$ and $y$.

We construct a grid of 3D model atmospheres with different horizontal variations of temperature. The atomic volume density in all the models is exponentially stratified following the expression
\begin{equation}\label{eq:nbar}
\overline{N(z)}=N_0\exp{\left(-\frac{z-z_0}{\beta}\right)}.
\end{equation}
In all our calculations we set the density scale height
$\beta=75$\,km and $N_0=10^{12}\,{\rm cm}^{-3}$. The vertical extension of the atmosphere $D_z = 2000\,\mbox{km}$ and
we set $z_0=0\,\mbox{km}$. The structure and adopted
values
are chosen to represent a
Sun-like
stationary atmosphere, while
the
distribution (\ref{eq:nbar}) corresponds to stratification established by gravitational and buoyancy forces.

In order to mimic the temperature variation of the inhomogeneous stellar
atmosphere we impose sinusoidal variation of kinetic temperature in the horizontal directions,
\begin{equation}
  T(x,y)=T_0\left[1+\alpha_T\sin{\left({}2\pi\frac{kx}{D_x}\right)}\sin{\left({}2\pi\frac{ky}{D_y}\right)}\right].\label{eqn:ktemp}
\end{equation}
This means that the mean temperature at any height $z$ (including the very bottom layer at $z_0$, where the perturbation in temperature is also assumed) in a model atmosphere is equal to $T_0$. We assume periodic boundary conditions in the horizontal directions of the model. The dimensionless parameter $\alpha_T\in\langle 0,1\rangle$
quantifies the amplitude of the temperature perturbation: The case of $\alpha_T=0$ corresponds to the 1D isothermal atmosphere while in the case $\alpha_T=1$ the perturbation is maximum. The dimensionless parameter $k$ determines the spatial period of the perturbation. For the sake of simplicity, we assume the same periods along the $x$ and $y$ axes. For numerical convenience, the horizontal dimensions of the model mesh is always set so that it is equal to one period of the fluctuation.
The actual domain size in km therefore depends on $k$. In the $k=1$ case, the horizontal domain size is $D_x\times D_y=1000\,\mbox{km}\times 1000\,\mbox{km}$. {For convenience, the spatial dimensions are hereafter expressed in Mm $=10^6$ m}. In all our calculations, we use $T_0=6000$\,K.

Similarly to Eq. (\ref{eqn:ktemp}), one could also introduce horizontal inhomogeneities in
density into the models (considering them separately, or simultaneously in phase or anti-phase), and study
their
impact on the emergent radiation. We
did
so for the
purpose
of a wider parametric study of the problem of inhomogeneous 3D atmospheres. However,
we found that
the effect of horizontal density distribution on line opacity (and thus the height of formation) imposes negligible changes in the resulting intensity spectrum, compared to the effect of temperature inhomogeneities on source function.
Therefore we decided to concentrate on the effect of temperature
perturbations.

Numerically, the mesh of the model is a cartesian grid with $100\times100\times120$ points along the $x$, $y$, and $z$ axes, respectively. The numerical solution of the NLTE problem is found using the code PORTA \citep{porta}. 

PORTA is a computer program designed for solving the problem of generation and transfer of spectral line intensity and polarization in three-dimensional models of stellar atmospheres. The numerical methods of the solution implemented in PORTA are highly convergent even in complex physical environments with steep gradients of physical variables and anisotropic radiation. 
{The implemented 3D formal solver of radiative transfer equation uses monotonic B\'{e}zier interpolation of source function along radiation beams within
the
short-characteristic scheme.} Among the best advantages is its efficient parallelization strategy, allowing simulation of radiation transfer using reasonably realistic 3D models. The time needed to calculate {one of our NLTE models (i.e. a model atmosphere with a fixed thermodynamic structure, for which we solve the NLTE problem)} was variable depending on specific properties of the model, typical calculation time was 3 -- 5 days using 61 cores. {We have calculated several tens of models with various inhomogeneity amplitudes and periods, while searching for the conditions under which the emergent radiation is affected.}

After the self-consistent solution of the NLTE problem is found, we calculate the formal solution of the radiative transfer equation at each surface point of the model atmosphere. We thus obtain the resulting intensity at all wavelengths at each surface point $(x_i,y_j),\,i,j=1,2,\ldots,N$, where $N$ is the number of grid nodes per $x-$ and $y-$ axes. From that information, we can construct synthetic spectrum of the considered line.

The impact of horizontal inhomogeneities on the spectral line profile
diminishes in the limiting cases of $k\to 0$ and $k\to\infty$.
On the other hand, the most significant impact of the horizontal radiation transfer can be expected if the spatial period of the perturbation is comparable to the mean photon destruction path $\ell^*$, which a photon can travel (while being absorbed and re-emited) before it becomes effectively destroyed \citep[see, e.g.][Chapter 14.2]{SA3}.
In such case, regions of the atmosphere with different thermal properties can interact radiatively. As we show below, the line-of-sight (LOS) effects can also play important role {in depth of formation of} the near wings of the line if $k\approx 1$ (discussed in Section \ref{ch:kpar}). In our models, we study the sensitivity of the line profile for $k\in[0.1,10]$.

We intentionally try to keep the models as simple as possible, in order to study solely the impact of inhomogeneous thermal structure of an atmosphere on the emergent radiation. Horizontal temperature inhomogeneities introduced by Eq. (\ref{eqn:ktemp}) are supposed to represent a physical asymmetry in thermodynamic structure of an atmosphere, which can be caused by, e.g., convective motions,
stellar
oscillation modes or MHD driven phenomena, as long as the necessary conditions are met (for instance the presence of surface magnetic field, macroscopic motions and/or any geometrical asymmetry might have to be taken into account). However, our models are purely academic, and are aimed to demonstrate the diagnostic
power
of 3D radiation transfer when studying such effects. More realistic models should be implemented for a direct application in stellar astrophysics.\footnote{Our calculations with PORTA also show that thermodynamic inhomogeneities significantly affect linear polarization of the emergent radiation. Considering the full Stokes vector (intensity, linear and circular polarization) can open more diagnostic possibilities. Polarization phenomena
are,
however, beyond the the scope of the
present
study.}


\section{Intensity of the emergent radiation}\label{ch:transfer}
In the following sections, we study the impact of temperature inhomogeneities on the
emergent radiation intensity spectrum. We intend to (a) find differences with respect to the unperturbed  model; (b) investigate the impact of 1D approximation on determination of the thermal stratification of the atmosphere.
Since the 1D stellar atmosphere usually serves as an approximate
representation instead
of a generally more realistic 3D case (where the solution of the NLTE problem is significantly more time-costly and technically difficult), we want to stress possible limitations of such simplification.


\subsection{Spatially resolved 3D case}\label{ch:results}
\begin{figure}[!htb]
   \centering
   \includegraphics[width=\hsize]{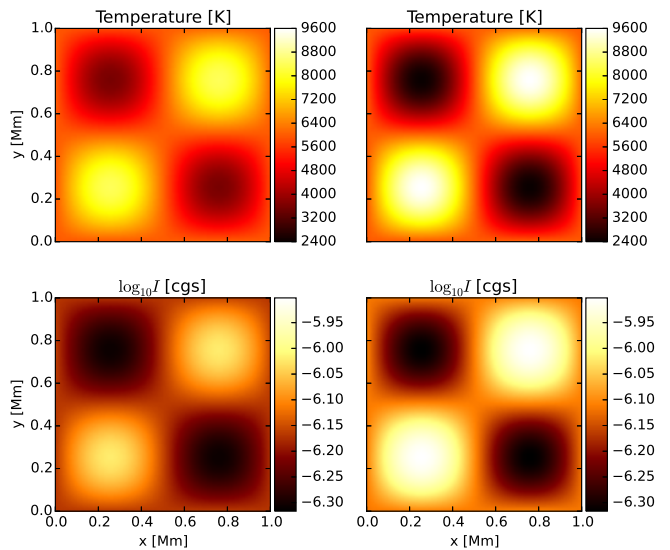}
   \caption{Top panels: variation of kinetic temperature for $\alpha_T=0.4$
(left panel) and $\alpha_T=0.6$ (right panel). The bottom panels show the corresponding line-center intensities in the direction along the $z$ axis, i.e., $\mu=\cos\theta=1$ (see Fig.~\ref{fig:sphere}).
}\label{fig:dt_baseT}%
   \end{figure}

\begin{figure}[!htb]
   \centering
   \includegraphics[width=.8\hsize]{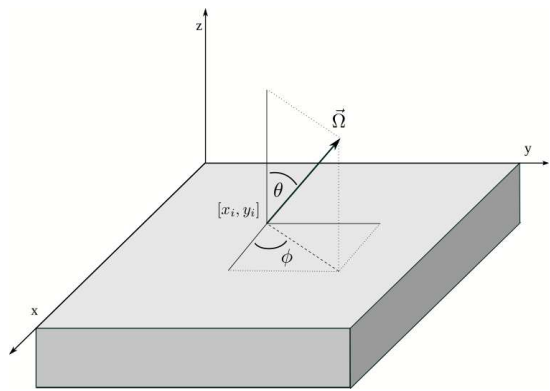}
   \caption{Definition of angles ${\theta}$ and $\phi$ of the line-of-sight vector $\vec\Omega$ in the cartesian coordinate system of the model.}
\label{fig:sphere}%
   \end{figure}

Horizontal thermal inhomogeneities defined in Eq.~(\ref{eqn:ktemp}) lead to horizontal variation of the emergent line intensity. In Fig. \ref{fig:dt_baseT} we show the variation of the temperature and of the corresponding {line-center emergent intensity in the direction $\mu=1$} in case of two model atmospheres. For definition
of the line-of-sight vector $\vec{\Omega}=(\theta,\phi)$
and the angles $\theta$, $\phi$ see Fig.\,\ref{fig:sphere}.
For the cosine of the angle $\theta$ we use the standard notation $\mu =
\cos\theta$.

Since the Planck function depends non-linearly on temperature, it follows that the mean Planck function at the bottom of the atmosphere, which is equal to the far-wing line intensity, depends on the perturbation amplitude $\alpha_T$. As we show below, similar effect can be found in a case of variation of the line source function at different heights of the atmosphere, which can lead to significant modification of the emergent line profile.

\subsection{Spatially integrated 3D case}\label{ch:emis}
The specific intensity of the emergent radiation $I(\vec{\Omega},\lambda,x_i,y_j)$ is then azimuthally and spatially averaged. By doing that, we make sure that the resulting synthetic spectra contain information about horizontal inhomogeneities, which are assumed to be spread across a stellar surface. The signal
at each surface grid point $(x_i,y_j)$ is integrated over azimuth $\phi\in\langle 0,2\pi\rangle$ by performing integration (using equidistant trapezoidal rule): 
\begin{equation}\label{eq:azimuth_int}
I(\mu,\lambda,x_i,y_j)=\frac{1}{2\pi}\int_0^{2\pi}\,\der\phi\,I(\vec{\Omega},\lambda,x_i,y_j).
\end{equation}
Thanks to the equidistant grid, the integration over the surface can be
replaced by a sum, and the resulting spatially integrated synthetic spectrum is finally given as an average over all surface points $(x_i,y_j)$, 
\begin{equation}\label{eq:synth_int}
I(\mu,\lambda)=\frac{1}{N^2}\,\sum_{i,j}\,I(\mu,\lambda,x_i,y_j) 
\end{equation}

{Radiative transfer is solved for each considered frequency, thus the procedure of obtaining synthetic spectra via expressions (\ref{eq:azimuth_int}) and (\ref{eq:synth_int}) is applied for continuum intensity value as well. Despite the presence of inhomogeneities, the atmosphere is for radiation at frequencies far enough from line-center completely transparent. Therefore the continuum intensity $I_{\rm C}$ is given by lower boundary condition and is identical to the Planck function value, averaged over the very bottom layer $z_0$, i.e., $I_{\rm C}=\langle B(T(x,y,z_0))\rangle_{z,y}$.}

\begin{figure}[!htb]
   \centering
   \includegraphics[width=.8\hsize]{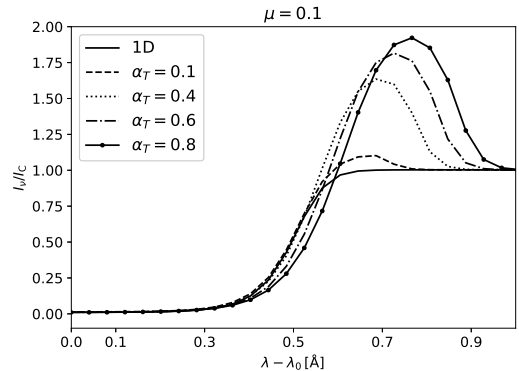}
   \caption{
Influence of different temperature perturbation amplitudes $\alpha_T$ on
spatially and azimuthally integrated emergent intensities normalized to
their respective continuum values $I_{\rm C}$. Intensities are plotted for the inclination $\mu=0.1$.
The solid line corresponds to the isothermal plane-parallel solution.}\label{fig:tintavr}%
   \end{figure}

Figure \ref{fig:tintavr} shows the azimuthally {and spatially} averaged specific intensity profiles $I(\lambda)$ computed for 3D models with inhomogeneity amplitudes $\alpha_T\in\{0.1,0.4,0.6,0.8\}$ and periodicity parameter $k=1$ (see Eq. \ref{eqn:ktemp}), as observed at inclination $\mu=0.1$ (close to the limb). Solid line corresponds to the case of an unperturbed (isothermal) atmosphere in planar geometry.

\begin{figure}[!htb]
   \centering
   \includegraphics[width=.8\hsize]{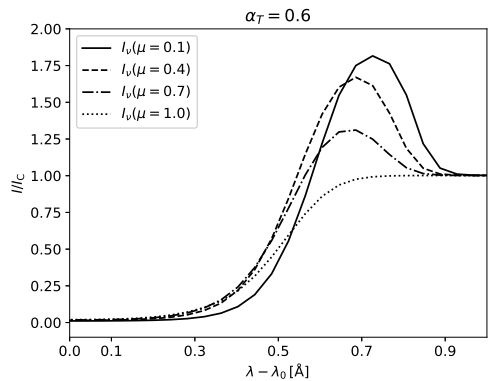}
   \caption{Emission peaks in synthetic intensity spectra appearing in solutions of inhomogeneous 3D models with temperature perturbation amplitude $\alpha_T=0.6$ for various inclination angles. The signal has been spatially and azimuthally averaged.}\label{fig:earsMu}%
   \end{figure}

Figure \ref{fig:earsMu} shows the inclination dependence of the azimuthally averaged specific intensity profiles $I(\lambda)$ computed for 3D model with inhomogeneity amplitude $\alpha_T=0.6$. The peaks decrease with increasing $\mu$ (having its maximum at $\mu=0.1$) and completely vanish for disc-center observation ($\mu=1.0$). Intensities in Figures \ref{fig:tintavr} and \ref{fig:earsMu} are normalized to continuum values $I_{\rm C}$ in order to emphasize the differences in wings.

In many physically relevant cases, the emergent line intensity can be
estimated using the Ed\-dington-Barbier relation \citep[e.g.,][]{SA3}, $I(\lambda,\mu)\approx S(\lambda,\tau_{\lambda}=\mu)$, where $\tau_{\lambda}$ is optical depth measured along the direction of beam propagation. The emergent radiation intensity is from atmospheric layers where the optical depth for a given wavelength and line-of-sight is around unity. The $\tau_{\lambda}=\mu$ surface is approximately the formation region of radiation at wavelength $\lambda$, and the Eddington-Barbier relation serves as a connection between the spectral features of line radiation and the locally calculated source function values.

Line profiles
plotted in Figure \ref{fig:earsMu} show that the
center-to-limb variation is affected by temperature inhomogeneities as well. Observing the corrugated $\tau_{\lambda}=\mu$ surface close to
the
limb ($\mu$ close to zero) means that certain areas of the surface become hidden from the line of sight, and the net intensity from visible areas may exceed the continuum value at certain wavelengths. This effect completely vanishes for observation at $\mu=1$, since the pattern of horizontal inhomogeneities is symmetric and, on average, the contribution of rays emerging from brighter areas is balanced by rays from dimmer areas.

However, if the variation of the source function along the line of sight is significant around $\tau=1$, the assumption of linearity of the source function with respect to optical depth no longer holds. Therefore, the
Eddington-Barbier relation cannot be reliably used to predict spectral features of the emergent line radiation, and the radiative transfer solution is required. In the following, we shall explain the connection of inhomogeneities with the calculated spectra, with special emphasis on the emission peaks appearing in almost all cases of inhomogeneous 3D models.

\begin{figure*}[!htb]
   \centering
   \includegraphics[width=\textwidth]{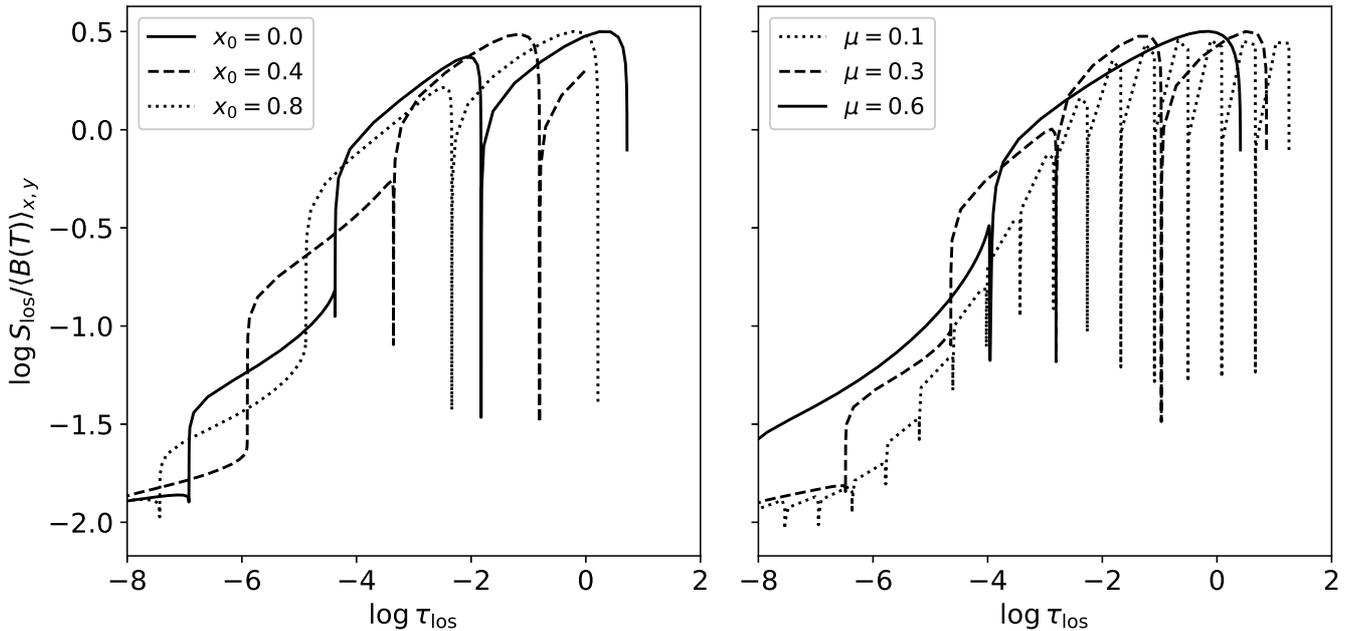}
   \caption{Source function normalized to the continuum value of intensity $\langle B(T)\rangle_{x,y}$ for various beams crossing the model surface at $y=0.25\,\mbox{Mm}$, namely at different points at a beam inclination $\mu=0.4$ (left panel), and at the point $x_0=0.0\,{\rm Mm}$ with various beam inclinations (right panel). Calculated for the model with temperature perturbation amplitude $\alpha_T=0.6$, the optical thickness along the line-of-sight is calculated at $\Delta\lambda=0.7\,$\AA.}\label{fig:s_los3beams}%
   \end{figure*}
  
The emission peaks at $\lambda-\lambda_0 = \Delta\lambda\approx 0.7\,$\AA{} may appear only if there is a number of rays originating in regions with the source function value higher than the Planck function averaged over the horizontal domain, i.e. $S/\langle B\rangle_{x,y}>1$ at $\tau_{\Delta\lambda=0.7}\approx\mu$. Figure \ref{fig:s_los3beams} illustrates this and shows the source function as a function of optical thickness {$\tau_{\rm los}$} along various beams {(line-of-sight)} crossing the surface of the model at $y=0.25\,{\rm Mm}$ at three different surface points $x_0=\{0.0,0.4,0.8\}\,{\rm Mm}$ with the same ray inclination $\mu=0.4$ (left panel), and beams crossing the surface at the same point $x_0=0.0\,{\rm Mm}$ with three various beam inclinations $\mu=\{0.1,0.3,0.6\}$ (right panel). The
half-width of the absorption profile is proportional to
the
square root of temperature; regions with lower temperature are less opaque to the radiation at
a
given frequency. When plotted as a function of
the {\em optical thickness} along the propagation beam,
the atmospheric regions with lower temperature shrink into optically thin areas with very low values of $S/\langle B\rangle$.
This causes sharp troughs in the plot.

\begin{figure*}[!htb]
   \centering
   \includegraphics[width=\textwidth]{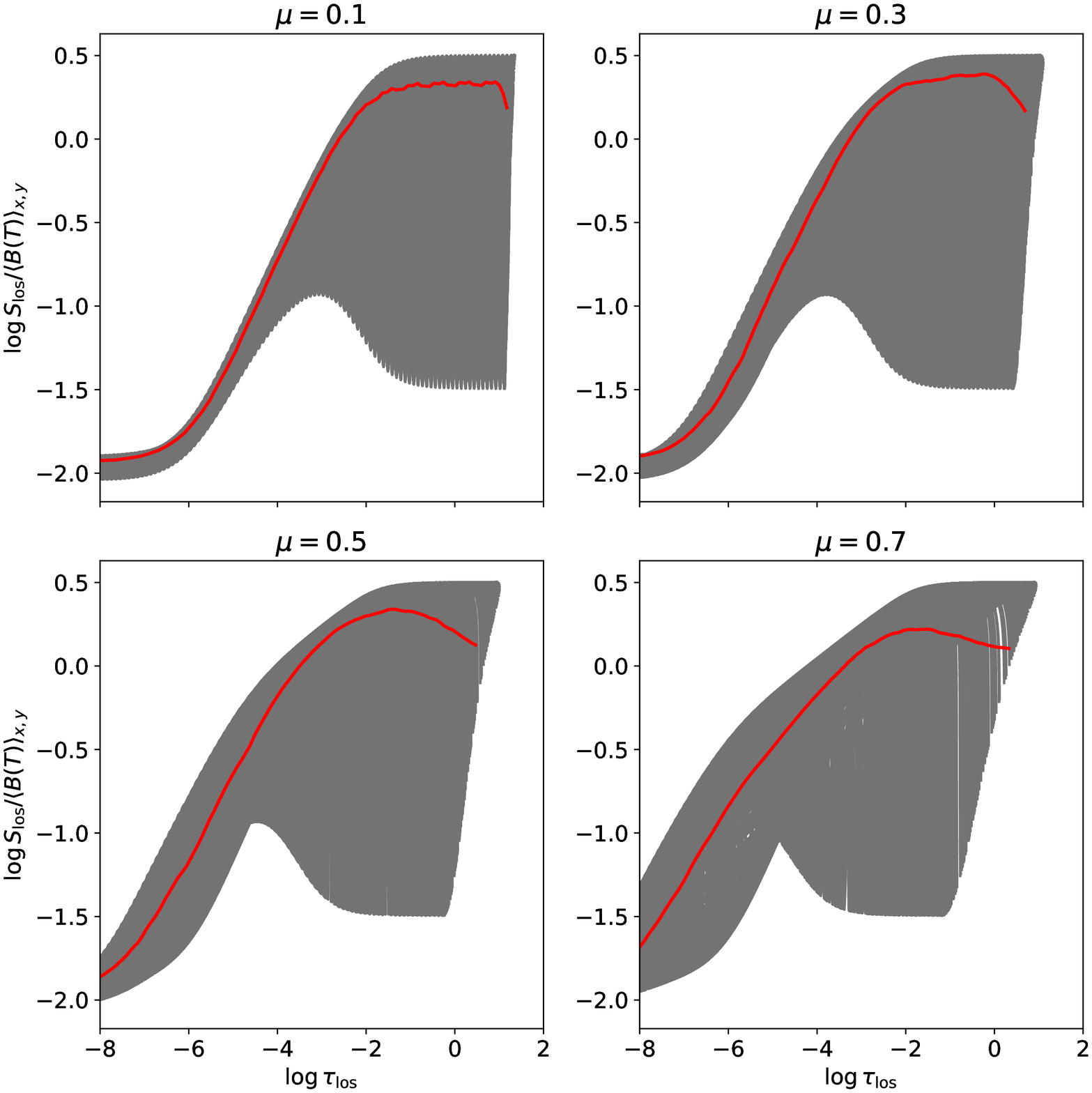}
   \caption{Source function normalized to the continuum value of intensity $\langle B(T)\rangle_{x,y}$ for 400 equally inclined rays
for four different inclinations crossing the model surface at different
random points $X_0$, calculated  for the model with temperature
perturbation amplitude $\alpha_T=0.6$.
The gray area approximately corresponds to possible values of source function along the given line-of-sight. Red line corresponds to 
the averaged value from all plotted rays. Top panels: inclination $\mu=0.1$ (left) and $\mu=0.3$ (right); bottom panels: inclination $\mu=0.5$ (left) and $\mu=0.7$ (right).
}\label{fig:s_los}
\end{figure*}

Because of periodic horizontal boundary conditions, each beam in Fig.~\ref{fig:s_los3beams} crosses the horizontal domain of the model several times according to the inclination angle, until it reaches the very bottom of the model. Figure~\ref{fig:s_los} shows the source function as a function of optical thickness along the line-of-sight, computed at wavelength close to emission peaks in line-wings ($\Delta\lambda\approx{}0.7\,$\AA) for 400 equally inclined beams crossing the surface of the perturbed model at random points
and for four different inclinations $\mu$.

The red lines in Fig.~\ref{fig:s_los} correspond to the average over all beams, indicating that there are always positive contributions to the source function close to $\tau=0.1$, giving higher average value of intensity than continuum. The gray areas roughly correspond to all possible values of source function along the line-of-sight. 

\section{Radiation flux from a spherical star}\label{ch:inte}
\begin{figure*}[!hbt]
   \centering
   \includegraphics[width=.7\linewidth]{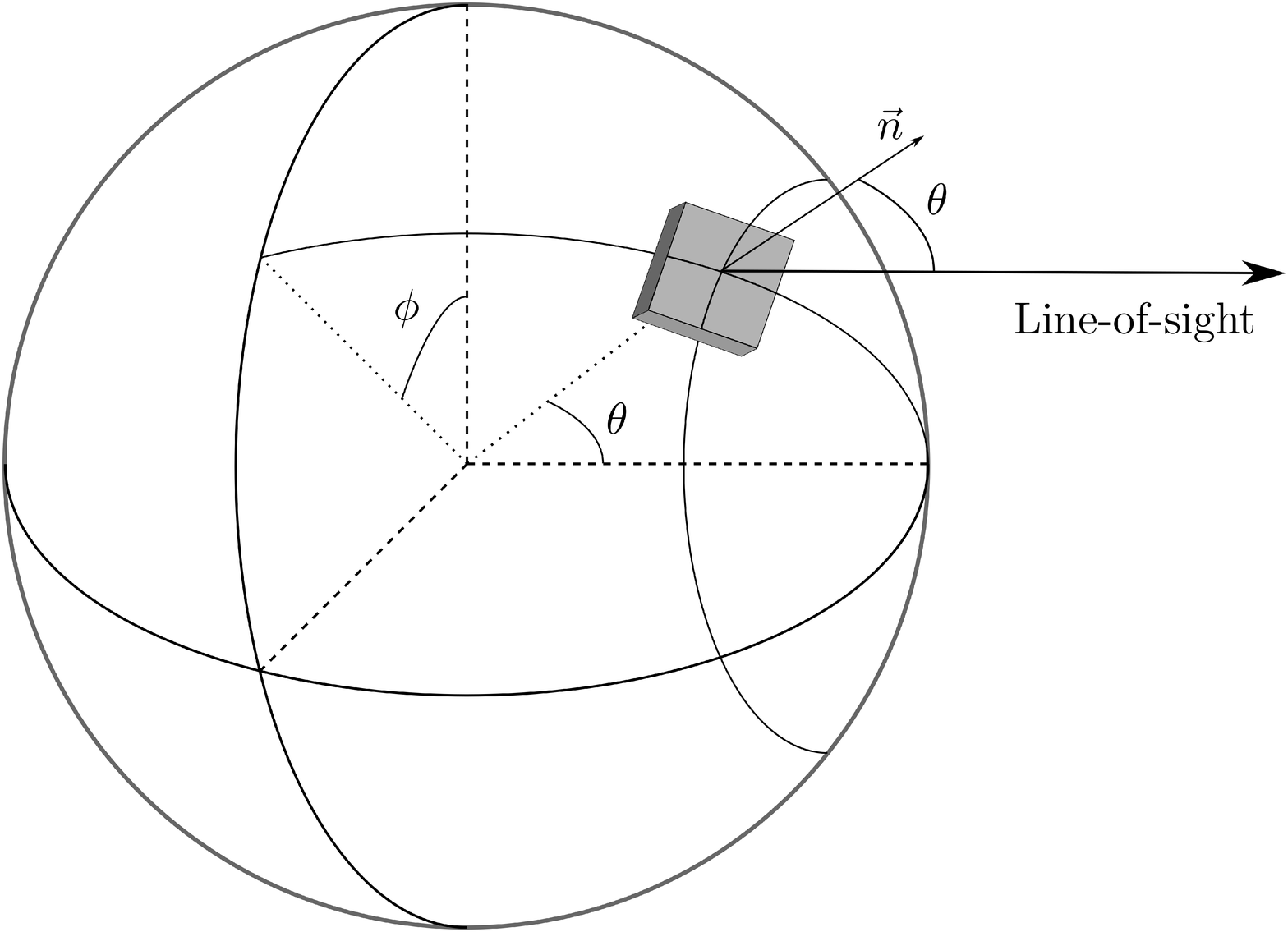}
   \caption{Geometry and definition of $\mu=\cos{\theta}$ and $\phi$ (not to scale, only sketch). The gray block illustrates the model atmosphere calculated using a 3D Cartesian grid. The stellar surface is observed from the direction of the line-of-sight, $\vec{n}$ indicates {the normal vector} to the surface of the considered model. Azimuth $\phi$ is measured in
the plane perpendicular to the line of sight.}\label{fig:StarSphere}%
   \end{figure*}

Given the lack of spatial resolution for stellar observations, we want to investigate how various temperature inhomogeneities affect the overall radiation flux emerging from a distant star, as measured by a distant observer. In the case of a non-irradiated spherical star, the emergent monochromatic specific radiation intensity integrated over the apparent stellar disk can be expressed using the monochromatic flux
\citep[see, e.g.,][]{unsold} as 
\begin{equation}\label{eq:surfI}
F_{\lambda}=\frac{1}{\pi}\int_0^{2\pi}\!\der\phi \int_0^1\! \der\mu\, \mu\, I_{\lambda}(\phi,\mu),
\end{equation}
where
$I_\lambda(\phi,\mu)$ {depends on both $\mu$ and $\phi$}.
The geometrical meaning of integration angles $\phi$ and $\theta$ (or equivalently $\phi$ and $\mu=\cos{\theta}$) is illustrated in Figure \ref{fig:StarSphere}; we identify $\phi$ and $\theta$ with the {polar angle} and inclination of models calculated on the Cartesian grid as indicated in Fig. \ref{fig:sphere}. 

Since we calculate emergent radiation from a spherically symmetric star
from planar atmospheric models, the geometry difference between planar
and spherical geometry (i.e. the curvature of the spherical atmosphere)
has to be taken into account. The principal difference between planar and spherical geometry is in treatment of radiation close to the limb. In planar geometry, all rays regardless the {polar angle} $\theta$ reach the lower boundary of the atmosphere, consequently the optically thick parts of the atmosphere. In spherical geometry, the almost tangent rays do not reach the optically thick part of the atmosphere and they emerge back at the upper boundary. As a consequence, the emergent radiation from the planar geometry is a bit overestimated
for rays close to $\mu=0$ \citep[see][]{KorcakovaEtAl}. To correct this, we start the angle integration at $\mu=0.1$ instead of $\mu=0$.
This way we assume that rays with the inclination in the interval
$\mu\in\langle0;1)$ go back to infinity and do not reach the optically
thick part of the atmosphere.
This corresponds to the estimate that the thickness of the atmosphere is $0.5\%$ of the stellar radius.

The integration is done in two steps. First we integrate the signal at each inclination $\mu$ over all possible azimuths $\phi\in\langle 0,2\pi \rangle$, then the integration over $\mu\in \langle 0.1,1 \rangle$ itself is performed (see the definition of geometry in Fig.~\ref{fig:StarSphere}). Integration over both azimuths (8 equidistantly separated directions in $\phi$) and inclinations (37 equidistantly separated directions in $\mu$) is done using the trapezoidal rule. 

\subsection{Influence of the inhomogeneity amplitude on line profiles}
\begin{figure}[!htb]
   \centering
   \includegraphics[width=.8\hsize]{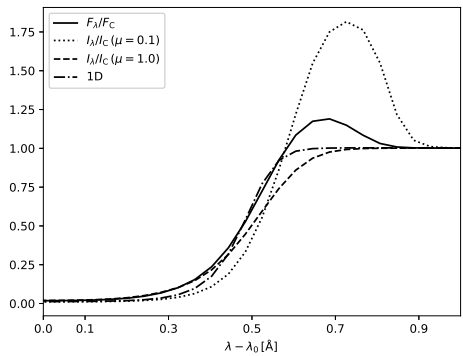}
      \caption{Intensity and flux profiles calculated for the 3D atmospheric model with horizontal temperature perturbation amplitude $\alpha_T=0.6$. Solid line corresponds to the signal integrated over the stellar surface using Eq.~(\ref{eq:surfI}); dashed and dotted lines correspond to the intensity of the 
emergent radiation from the same model, observed at inclinations $\mu=1.0$ and $\mu=0.1$, respectively.}
         \label{fig:one}
   \end{figure}

   Fig. \ref{fig:one} shows the intensity profile of the radiation emerging from the horizontally perturbed 3D model with temperature inhomogeneity amplitude $\alpha_T=0.6$ as observed close to the limb ($\mu=0.1$) and at the disc-center ($\mu=1.0$). Solid line corresponds to {the} solution of the same model, but integrated over all azimuths $\phi$ and inclinations $\cos{\theta}\equiv\mu$ as indicated by Eq. (\ref{eq:surfI}). Emission peaks at line wings are strongest in the case of limb observations, completely vanishing at the disc center. Interestingly, the emission still remains even after averaging over the stellar disc, which is a direct consequence of temperature inhomogeneities in atmospheric {structure, mainly in its lower regions}.

\begin{figure}[!htb]
   \centering
   \includegraphics[width=.8\hsize]{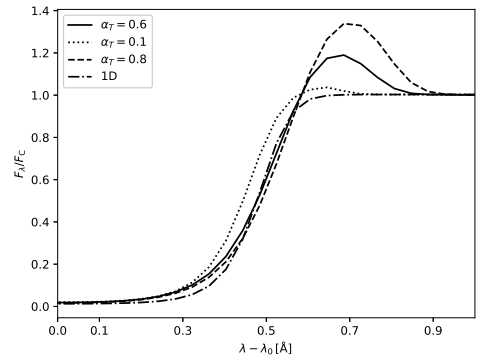}
      \caption{Profiles of the emergent flux calculated by Eq. (\ref{eq:surfI}) for horizontally inhomogeneous 3D atmosphere models with perturbation amplitudes $\alpha_T=0.1$ (dotted line), $\alpha_T=0.6$
(full line) and $\alpha_T=0.8$  (dashed line).}
         \label{fig:two}
   \end{figure}

Fig. \ref{fig:two} shows {the line} flux calculated using Eq. (\ref{eq:surfI}) for inhomogeneous 3D models with different perturbation amplitudes $\alpha_T=0.1$, $\alpha_T=0.6$, and $\alpha_T=0.8$. {The} shape of the spectral line is, as expected, behaving similarly to {the} intensities shown in Fig.~\ref{fig:tintavr} in accordance with the value of the amplitude $\alpha_T$ -- the overall flux increases nonlinearly with perturbation amplitude. It is due to the strong and nonlinear temperature-dependence of the Planck function, which is coupled to the source function at $\tau_{\lambda}=1$. The {largest differences} in flux magnitude among the plotted models are in the line-core (formed in the upper part of the model atmosphere) and at line-wings (formed at mid- to lower model's parts). Emission peaks seems to almost vanish for the model with the smallest perturbation amplitude $\alpha_T=0.1$, as the temperature-driven effect is by angle-averaging for such small inhomogeneity suppressed.

\subsection{Influence of the perturbation period on line profiles}\label{ch:kpar}
\begin{figure}[!tbh]
   \centering
   \includegraphics[width=.8\hsize]{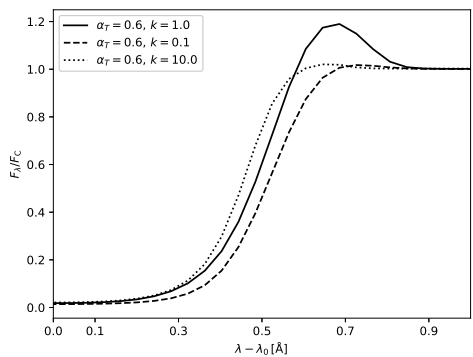}
      \caption{Emergent radiation flux calculated via Eq. (\ref{eq:surfI}) for horizontally inhomogeneous 3D atmosphere models with temperature perturbation amplitude $\alpha_T=0.6$; plotted are fluxes for models with various perturbation periods: $k=0.1$ (solid line), $k=1.0$ (dashed line) and $k=10.0$ (dotted line). The emission peaks remain dominant only for the model with a single perturbation period along the horizontal domain.}
         \label{fig:fou}
   \end{figure}

The periodicity parameter $k$ has an important impact on "visibility" of temperature inhomogeneities in calculated spectra, because it directly corresponds to its spatial scales with respect to photon destruction path $\ell^*$. Fig.~\ref{fig:fou} shows results of 3D line profile calculations for three different values of
the parameter $k$, all for the perturbation amplitude $\alpha_T=0.6$, namely for $k=0.1$, $1.0$, and $10.0$.
     
In the limit of low perturbation period ($k=0.1$), the variation of temperature takes place on 10 {times} larger spatial scales, such that the influence of inhomogeneities on horizontal transfer of radiation becomes negligible, making the problem effectively 1.5D. Thus, we expect the emission effect at $\Delta\lambda\approx{}0.7$ of the line to disappear in the spectrum of the radiation flux. On {the} contrary, in the limit of high perturbation period ($k=10$), the temperature fluctuates on 10 {times} smaller spatial scales. Horizontal transfer of radiation effectively reduces the impact of individual inhomogeneities, causing the emission in calculated spectra to disappear.

\subsection{Comparison of 1D and 3D solutions}
    Temperature variation along the horizontal direction of a stellar atmosphere can leave certain imprints upon the surface-integrated emergent radiation flux. However, the question about the unambiguity of the interpretation of such spectra is crucial, especially if it is used to determine the thermodynamic structure of an observed stellar atmosphere. If
there exists a different atmospheric model which produces similar emergent line profiles, we are not be able to decide which model better represents the atmospheric structure. 

\begin{figure}[!htb]
   \centering
   \includegraphics[width=.6\hsize]{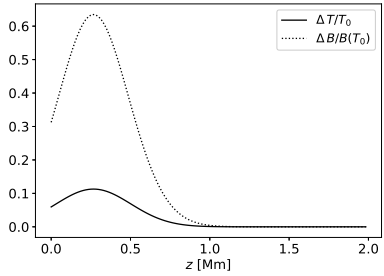}
      \caption{Vertical temperature stratification of the plane-parallel atmosphere, given by Eq. (\ref{eq:tezet}) with $z_c=16$, $2w^2=400$, $f=0.11$ and the reference temperature $T_0=6000\,{\rm K}$;
relative change in temperature with height in the atmosphere (solid line). Dotted line corresponds to related relative change of the Planck function.}
         \label{fig:str}
   \end{figure}

It is indeed possible to find solutions which give, to some extent, very similar spectral shape of the
emergent radiation flux.
Let us
modify the vertical thermal structure of a plane-parallel atmosphere as
a Gaussian perturbation along the $z$-axis of the model,
\begin{equation}\label{eq:tezet}
T(z)=T_0\left[1+f\exp{\left(-\frac{(z-z_{\rm c})^2}{2w^2}\right)}\right],
\end{equation}
where $z_{\rm c}$ is the geometrical depth of the temperature maximum, $w$ is a parameter influencing the width of the Gaussian profile and $f$~is an auxiliary numerical factor reducing the maximal allowed change in temperature.
Figure \ref{fig:str} shows an example of a vertical temperature stratification according to Eq. (\ref{eq:tezet}) and corresponding Planck function variation, both in relative units.
The Planck function directly affects the source function at given height. The set of free parameters of the temperature structure in Fig.~\ref{fig:str} was chosen intentionally to reproduce similar emission effects as we obtain from the horizontally inhomogeneous 3D models described in Section~\ref{ch:emis}.

We used the
trial and error strategy
to obtain similar shape of the resulting line profile (i.e. emission
peaks in line wings) by changing four free parameters $z_c$, $w$, $f$ and
$T_0$ in Eq. \eqref{eq:tezet}.
For a given set of the free parameters we solve the NLTE problem.
Then we calculate flux and compare it to the flux emerging from an inhomogeneous 3D model in consideration. While changing one parameter at a time, we figure out how each of them affects the solution.

\begin{figure}[!hbt]
   \centering
   \includegraphics[width=.8\hsize]{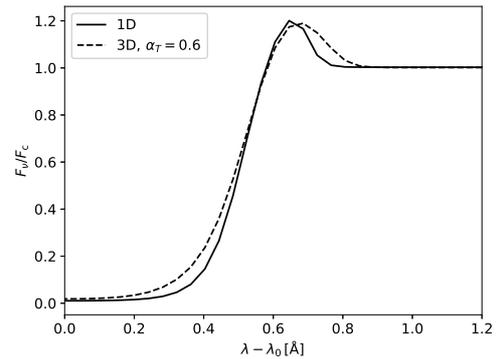}
      \caption{
Radiation flux profile calculated for the horizontally inhomogeneous 3D atmosphere model with temperature perturbation amplitude $\alpha_T=0.6$ (dashed line), obtained by applying Eq. (\ref{eq:surfI}). Solid line corresponds to the flux calculated for the plane-parallel atmosphere model, which gives similar spectral shape of the radiation flux. Its vertical temperature structure is given by Eq. (\ref{eq:tezet}) with $z_0=16$, $2w^2=400$, $T_0=6000\,$K and $f=0.11$.}
         \label{fig:thr}
   \end{figure}

We have found several models with a vertically dependent temperature structure giving similar emission peaks in line-wings as the model with horizontal temperature perturbation, but with various differences across the wavelength range. Fig.~\ref{fig:thr} shows
a line profile emerging from such vertically perturbed atmosphere, which
is most similar to a profile from a horizontally inhomogeneous 3D
atmosphere, equivalent in its emission maximum and slightly {differing}
in a shape.
These two similar line profiles forming in very different
physical conditions serve as an example of ambiguity of emergent
radiation, which may appear for stars of any spectral type.

\section{Conclusions}\label{ch:conclusion}
The temperature structure of a stellar atmosphere has a significant impact on the intensity of spectral lines, which are formed at inhomogeneous regions. This is due to the presence of areas with higher values of the local source function, which tend to create higher values of intensity of a spectral line at corresponding wavelengths. This effect can influence spectra emerging from the stellar surface.

As a consequence, if we try to interpret the observed spectra using 1D atmosphere models, we can make erroneous conclusions about the atmospheric structure. A real stellar atmosphere (its spatial distribution of temperature, atomic volume density, etc.) is naturally very complicated, so it is to be expected that perturbations of various origin and geometrical shape are quite common. In such a case, the 
multidimensional approach to the NLTE problem can provide a better tool to determine the atmospheric structure. However, it strongly depends on the nature of inhomogeneities -- its shape, size and spatial distribution over the stellar surface.

 
\section*{Acknowledgements}
The computations were performed on the parallel computer cluster OCAS
at the Astronomical Institute Ond\v{r}ejov
We thank Dr.~J.~\v{S}t\v{e}p\'an for the possibility to use the code
PORTA and for many invaluable discussions.
This research was supported by a grant GA\,\v{C}R 16-01116S.
The Astronomical Institute Ond\v{r}ejov is supported by a project
\mbox{RVO:67985815} of the Academy of Sciences of the Czech Republic.
%

\end{document}